\documentclass[aps,prl, two column, superscriptaddress, letter]{revtex4}
\usepackage{graphicx}
\usepackage{textcomp}
\usepackage{color}
\usepackage{dcolumn}
\usepackage{multirow}
\usepackage{ulem}
\usepackage{textcomp}
\begin{document}

\title {State Transfer Between a Mechanical Oscillator and Microwave Fields in the Quantum Regime}

\author{T. A. Palomaki}
\affiliation{JILA, National Institute of Standards and Technology
and the University of Colorado, Boulder, CO 80309, USA} \affiliation{Department of
Physics, University of Colorado, Boulder, CO 80309, USA}

\author{J. W. Harlow}
\affiliation{JILA, National Institute of Standards and Technology
and the University of Colorado, Boulder, CO 80309, USA} \affiliation{Department of
Physics, University of Colorado, Boulder, CO 80309, USA}

\author{J. D. Teufel}
\affiliation{National Institute of Standards and Technology, Boulder, CO 80305, USA}

\author{R. W. Simmonds}
\affiliation{National Institute of Standards and Technology, Boulder, CO 80305, USA}

\author{K. W. Lehnert} \email{konrad.lehnert@jila.colorado.edu}
\affiliation{JILA, National
Institute of Standards and Technology and the University of
Colorado, Boulder, CO 80309, USA} \affiliation{Department of
Physics, University of Colorado, Boulder, CO 80309, USA}

\date{\today}

\begin{abstract}

Recently, macroscopic mechanical oscillators have been coaxed into a regime of quantum behavior, by direct refrigeration \cite{O'Connell2010} or a combination of refrigeration and laser-like cooling \cite{Teufel2011b, Chan2011}. This exciting result has encouraged notions that mechanical oscillators may perform useful functions in the processing of quantum information with superconducting circuits \cite{Hertzberg2010, Sillanpaa2009, LaHaye2009, REGAL2008, O'Connell2010}, either by serving as a quantum memory for the ephemeral state of a microwave field or by providing a quantum interface between otherwise incompatible systems \cite{Wang2012, Regal2011}. As yet, the transfer of an itinerant state or propagating mode of a microwave field to and from a mechanical oscillator has not been demonstrated owing to the inability to agilely turn on and off the interaction between microwave electricity and mechanical motion. Here we demonstrate that the state of an itinerant microwave field can be coherently transferred into, stored in, and retrieved from a mechanical oscillator with amplitudes at the single quanta level. Crucially, the time to capture and to retrieve the microwave state is shorter than the quantum state lifetime of the mechanical oscillator. In this quantum regime, the mechanical oscillator can both store and transduce quantum information.
\end{abstract}

\maketitle

Mechanical oscillators are particularly appealing for storing or transducing quantum information encoded in microwave fields, as their fabrication is compatible with, and their size similar to, superconducting quantum circuits. Indeed, a high-frequency mechanical oscillator has been combined with a superconducting qubit, demonstrating the transfer of a qubit state to the oscillator \cite{O'Connell2010}. While an impressive demonstration, the few nanosecond lifetime of the oscillator, which was much shorter than the lifetime of the qubit, suggests that lower frequency oscillators with much longer lifetimes may form superior quantum memories. This is particularly true for storing the information in an itinerant mode of a microwave field, for which the characteristic time to acquire a new value is about 100~ns \cite{Houck2007, Mallet2011, Eichler2011}. The penalty for working with a low-frequency oscillator is that the oscillator will not naturally be in its quantum ground state but rather a hot thermal state. For example, an oscillator with a resonance frequency of $\Omega_\mathrm{m}=10$~MHz in equilibrium with an environment at temperature $T_{\mathrm{env}}=25$~mK will contain a statistically fluctuating number of quanta with average value $n_{\mathrm{env}}\equiv [\mathrm{exp}(k_\mathrm{B} T_{\mathrm{env}}/ \hbar\Omega_\mathrm{m})-1]^{-1}=50$ quanta. Fortunately, embedding the mechanical oscillator in a high frequency resonant circuit creates an interaction between the mechanical oscillator and itinerant microwave fields that can be used cool the oscillator to its ground state \cite{Teufel2011b}.

Here we show that this same interaction exchanges a coherent state of an itinerant microwave field with the hot thermal state of the mechanical oscillator. This exchange simultaneously transfers the information from the itinerant microwave field to the mechanical oscillator and removes the thermal excitations from the mechanical oscillator, preparing it in a low-entropy state. We transfer coherent microwave fields with amplitudes at the single quanta level while removing all but one thermal excitation from the  mechanical oscillator. We demonstrate that this low entropy state is preserved for a characteristic time of 90~$\mu$s, about 125 times longer than the minimum transfer time. Finally, we contrast the transfer of itinerant states of a microwave field and localized states of the microwave resonant circuit in which the mechanical oscillator is embedded.

Our mechanical oscillator is the fundamental drumhead mode of a thin (100 nm), 15 $\mu$m diameter superconducting aluminum membrane, which forms the upper plate of a 50 nm vacuum gap capacitor (Fig. 1a) \cite{Cicak2010}. This arrangement forms a parallel plate capacitor that, together with a spiral inductor, creates our microwave resonant circuit with resonant frequency $\omega_\mathrm{c} \cong 2\pi \times 7.5$ GHz. We couple energy to and from the circuit via a nearby transmission line and determine the device parameters from spectroscopic measurements (similar to Ref. \cite{Teufel2011b}) in a dilution refrigerator with a 15 mK base temperature. The tension and diameter of the membrane produce a fundamental mode $\Omega_{\mathrm{m}}=2\pi \times 10.5$ MHz, with a linewidth $\gamma_\mathrm{m}=2\pi \times 35$ Hz. This mode and the oscillator mass $m=48$ pg imply a zero-point motion $x_{\mathrm{zp}}=4.1$ fm.

The key to our state transfer measurements is the ability to rapidly switch the coupling between the mechanical oscillator and the resonant circuit on and off \cite{JacobsK2011, Hofer2011, Romero2011, Machnes2012, Vanner2011}. That the two systems are coupled can be understood from the fact that the motion of the oscillator changes the circuit's resonance frequency $\omega_\mathrm{c}$. The coupling can be described by an interaction Hamiltonian $H_{\mathrm{int}}=\hbar G \hat{x} \hat{a}^\dagger \hat{a}$, where $G=d\omega_\mathrm{c}/dx$, $\hat{a}(\hat{a}^\dagger)$ corresponds to the annihilation (creation) operator for microwave photons, $(\hat{a}^\dagger \hat{a})$ is the photon number
and $\hat{x}$ is the position operator of the mechanical oscillator. In the presence of a strong microwave excitation at frequency $\omega_\mathrm{d}=\omega_\mathrm{c}-\Omega_\mathrm{m}$, the interaction between the two systems is greatly enhanced and takes the approximate form
\begin{equation}\label{eq:LinInteractHamil}
    H_{\mathrm{int}}=\hbar g_{\mathrm{0}} \sqrt{N_\mathrm{d} (t)}(\hat{a} \hat{b}^\dagger+ \hat{b} \hat{a}^\dagger)\;,
\end{equation}
where $N_\mathrm{d} (t)$ is the strength of the excitation at $\omega_\mathrm{d}$ expressed as the number of photons in the resonator, $g_\mathrm{0} = G x_\mathrm{zp}=2\pi \times 200$~Hz, and $\hat{b} (\hat{b}^\dagger)$ is the annihilation (creation) operator for mechanical phonons \cite{Groblacher2009a}. The interaction Hamiltonian now resembles that of a beam splitter and as such it coherently exchanges energy between microwave fields at $\omega_\mathrm{c}$ and mechanical phonons, as highlighted by the blue arrow in Fig. 1b. Crucially, the rate of that exchange -- the transmission of the beam splitter -- is controlled by $N_\mathrm{d}(t)$, a quantity that we can change. In addition to the dynamics contained within Eq.~\ref{eq:LinInteractHamil}, energy is also coupled between the resonant circuit and itinerant modes in the transmission line at a rate $\kappa_{\mathrm{ext}}= 2\pi \times 160$ kHz. The resonators linewidth is $\kappa=\kappa_{\mathrm{in}}+\kappa_{\mathrm{ext}}$, where $\kappa_{\mathrm{in}}$ is the undesired internal loss rate, a quantity that depends on the energy in the resonant circuit \cite{Gao2008a}. For the range of microwave powers used here we find $\kappa$ between $2\pi \times 220$ kHz and $2\pi \times 250$ kHz. Equation~\ref{eq:LinInteractHamil}, which permits the rapid and coherent exchange between electrical and mechanical energy, is an approximation that assumes $\kappa$ is small enough to satisfy $\kappa\ll\Omega_\mathrm{m}$ (the resolved sideband limit).

Far in the resolved sideband limit, the microwave fields incident on the circuit can be divided conceptually into two types of fields based on their frequency (Fig. 1c). Fields with frequency near $\omega_\mathrm{c}$ prepare the circuit (preparation field) in a particular state, which can be transferred into the mechanical oscillator. Fields with frequencies near $\omega_\mathrm{c}-\Omega_\mathrm{m}$ transfer states (transfer field) between the mechanical oscillator and the microwave circuit. In the resolved sideband limit, these fields can remain spectrally distinct even when their strengths are varied more rapidly than $\kappa$. The transfer field's strength (Fig. 1b) determines whether itinerant fields in a nearby transmission line or states of the resonator are transferred. Because the resonator's state evolves into an itinerant mode at rate $\kappa_{\mathrm{ext}}$, if $2g_\mathrm{0} \sqrt{N_\mathrm{d}} < \kappa$, the resonator never completely contains the former or future state of the mechanical oscillator and the transfer process is between itinerant microwave fields and the oscillator \cite{Fiore2011, Chang2011}. If instead the coupling rate is increased such that $2g_\mathrm{0} \sqrt{N_\mathrm{d}} > \kappa$ (strong coupling regime), the state of the resonator can be swapped with the oscillator \cite{Verhagen2012, Groblacher2009a}.

The phenomena of cooling, state measurement, and optomechanically induced transparency \cite{Verhagen2012} can be understood as steady state limits of the transfer process \cite{Teufel2011} in either regime. For example, if the preparation field is unexcited in its vacuum state, continuous application of the transfer field exchanges the thermal state of the mechanical oscillator for the ground state of the microwave field, cooling the oscillator \cite{Teufel2008a, Rocheleau2010, Marquardt2007, Wilson-Rae2007}. Microwave fields emitted by the circuit carry away the thermal state where it can then be measured. If instead, a transfer field and a preparation field are both applied continuously, they produce an interference effect in the response of the circuit to the preparation field (optomechanically induced transparency) \cite{Safavi2011, Teufel2011, Verhagen2012}. All of these phenomena are contained within the classical equations of motion for the coupled system; however, if the transfer rate between phonons and itinerant photons \begin{equation}\label{eq:Gamma}
    \Gamma_{\mathrm{ext}}=\frac{4 g_\mathrm{0}^2 N_\mathrm{d}}{\kappa} \frac{\kappa_{\mathrm{ext}}}{\kappa} \;,
\end{equation}
exceeds the rate at which a single phonon is exchanged with the oscillator's environment, the transfer process is capable of exchanging single mechanical phonons and microwave photons. The factor $\kappa_{\mathrm{ext}}/\kappa$ in Eq.~\ref{eq:Gamma} distinguishes the transfer rate to intinerant photons from the total decay rate to all channels $ \Gamma=4 g_\mathrm{0}^2 N_\mathrm{d} /\kappa$. A single phonon enters from the environment at a rate given by $\simeq n_\mathrm{env}\gamma_\mathrm{m}$, where $\gamma_\mathrm{m}$ is the rate at which the oscillator exchanges energy with the environment. The ratio of the state transfer rate to the environmental decoherence rate takes the from of a cooperativity parameter $C=4g_\mathrm{0}^2 N_\mathrm{d}/(\kappa n_\mathrm{env}\gamma_\mathrm{m})$, conveniently marking entry to the quantum regime when $C>1$. The device we study is well suited for demonstrating state transfer because it is far in the resolved sideband limit and capable of achieving $C\gg1$.

\begin{figure}
  \centering
  \includegraphics[width= 3 in]{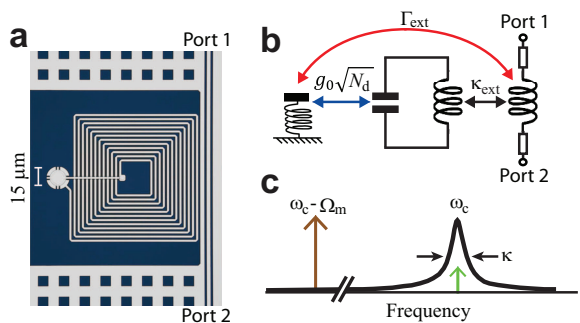}\\
  \caption{\textbf{Schematic description of the experiment. a,} False-colour image of the device; the sputtered aluminum is shown in grey and the sapphire substrate in blue. A coplanar-waveguide transmission line, visible at the right edge of the image couples itinerant microwave fields to the resonant circuit via mutual inductance with the spiral inductor of the resonator. The mechanical oscillator is the 15~$\mu$m disk visible at the left edge of the figure. \textbf{b}, The diagram represents the exchange of energy between itinerant microwave fields and mechanical motion. In the presence of a transfer field with $N_\mathrm{d}$ photons, the coupling rate between the resonator and the oscillator fields is $g_0 \sqrt{N_\mathrm{d}}$ and the resonator energy enters from and departs to the transmission line at a rate $\kappa_{\mathrm{ext}} $. The transfer rate between itinerant microwave fields and mechanical motion is $\Gamma_{\mathrm{ext}}$. The preparation and transfer field are applied at port~1 and fields that emerge from the circuit are measured at port~2 with a sensitive microwave receiver circuit. \textbf{c,} A frequency domain representation of the experiment shows that the preparation field (green) and the transfer field (brown) are widely seperated ($\Omega_\mathrm{m} \gg \kappa$) compared to the linewidth of the circuit's response (black).
}\label{fig:fig1}
\end{figure}

\begin{figure}
  \centering
  \includegraphics[width= 3 in]{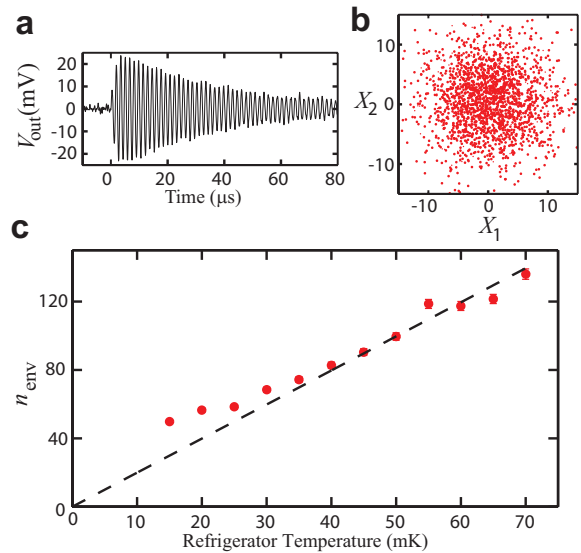}\\
  \caption{\textbf{Measurement by state transfer. a,} The mixed-down microwave signal $V_{\mathrm{out}}(t)$ is measured after the application of a transfer field at time $t=0$ for a case where the mechanical oscillator has been prepared in a state with large amplitude. \textbf{b,} Each point is an estimate of the state of the mechanical oscillator expressed as quadrature amplitudes $X_1$ and $X_2$ extracted from data similar to that shown in \textbf{a}. The points are 2000 independent measurements of the state of the mechanical oscillator when in equilibrium at $T_{\mathrm{env}}=25$~mK. \textbf{c,} The red points show the average energy $N$ of the oscillator (in units of quanta) as a function of the refrigerator's temperature, calculated from the measurements of the oscillator's state as $N = \langle (X_1^2 + X_2^2) \rangle - 1/\eta$. By assuming the mechanical oscillator is in equilibrium with an environment we measure $n_{\mathrm{env}}=N$, which grows with the refrigerator's temperature and scales linearly (dashed line) with temperature between 35 mK and 70 mK.
  }\label{fig:fig2}
\end{figure}

We first use our ability to transfer the state of a mechanical oscillator to an itinerant microwave field in order to measure the state of the oscillator. Specifically, we energize the transfer field at time $t=0$, with $\Gamma > n_{\mathrm{env}}\gamma_\mathrm{m}$ sufficiently large to transfer the state out of the mechanical oscillator into a microwave signal propagating in the transmission line before the oscillator can absorb one quantum from its warm environment. To recover the state of the mechanical oscillator, we amplify, mix down, and digitize the microwave signal. The frequency component at $\omega_\mathrm{d} + \Omega_\mathrm{m}$ in the microwave field is mapped to oscillations in $V_{\mathrm{out}}$ at frequency $\omega_{\mathrm{out}}/2\pi \approx1$~MHz. The mixed-down signal $V_{\mathrm{out}}$ encodes the amplitude and phase of the mechanical oscillator at $t=0$. (See Supplementary information.) To illustrate this procedure, in Fig.~\ref{fig:fig2}a we show such a measurement; $V_{\mathrm{out}}(t)$ is plotted for a case where the mechanical oscillator was first prepared in a state of large amplitude so that the oscillations can easily be resolved. The oscillations decay exponentially with a rate $(\Gamma + \gamma_\mathrm{m})/2 \approx \Gamma/2 = 2\pi \times 4$ kHz, chosen by the magnitude of the transfer field $\sqrt{N_\mathrm{d}}$. Using our knowledge of $\Gamma$ and $\omega_{\mathrm{out}}$, we can optimally extract the amplitude and phase of the microwave temporal mode from any particular realization of the measurement, and thus infer the state of the oscillator just before it was transferred. Rather than work with amplitude and phase variables, we prefer to describe the state of the oscillator in terms of quadrature amplitudes $\hat{X_1}=(\hat{b}e^{i\Omega_\mathrm{m} t}+\hat{b}^{\dagger} e^{-i\Omega_\mathrm{m} t})/2$ and
$\hat{X_2}=(\hat{b}e^{i\Omega_\mathrm{m} t}-\hat{b}^{\dagger} e^{-i\Omega_\mathrm{m} t})/2i$, which have the virtue of being canonically conjugate. We refer to our inference of $\hat{X_1}$ and $\hat{X_2}$ in any single measurement as $X_1$ and $X_2$. This inference is not perfect because noise with both quantum and technical origins is added by the state transfer and subsequent measurement of the microwave field. We independently determine the variance of this measurement noise and denote it $1/\eta$, where $\eta \leq 1$ is an effective quantum efficiency of the measurement (See Supplementary Information).

To demonstrate that we are in fact measuring the state of a mechanical oscillator by transfer to an itinerant microwave field and to calibrate our measurement, we examine the statistics of our measurements with the mechanical oscillator prepared in a thermal state by equilibration with the environment. In our protocol, we wait ($\approx 5/\gamma_\mathrm{m}$) for the oscillator to come into equilibrium with the environment and then apply a transfer tone for a time longer than $5/\Gamma$, extracting a pair of values (as yet uncalibrated) $X_1$ and $X_2$. By repeating the same protocol 2000 times we gather sufficient statistics to determine the state of the mechanical oscillator. For a mechanical oscillator in equilibrium with an environment at temperature $T_{\mathrm{env}}$ we expect the measurements of $X_1$, $X_2$ to follow a two dimensional Gaussian distribution whose variance depends on the environment's temperature as $\langle X_1^2 + X_2^2 \rangle \propto n_{\mathrm{env}} + 1/\eta$, where the brackets denote an ensemble average. In Fig. 2b we plot the 2000 measurements, demonstrating that the measurements have a distribution consistent with a thermal state (see Supplementary Information). By raising the temperature of the dilution refrigerator we expect the variance to increase accordingly if the mechanical oscillator's environment is at the refrigerator's temperature. Indeed, between 35 and 70 mK, we find that $\langle X_1^2 + X_2^2 \rangle - 1/\eta$ is proportional to the refrigerator's temperature as shown in Fig. 2c. In this temperature range we conclude that the oscillator is in equilibrium with the refrigerator; therefore, we can calibrate $X_1$, $X_2$ and $1/\eta$ into dimensionless units with a single scaling factor such that $\langle X_1^2 + X_2^2 \rangle - 1/\eta = n_{\mathrm{env}}$. We consistently find $\eta$ between 0.11 and 0.12, close to the value we expect (see Supplementary Information). Below 35~mK the mechanical oscillator falls out of equilibrium with our refrigerator, cooling only to 25 mK when the refrigerator is at 15~mK. This saturation is not caused by the pulsed nature of our measurement as it is also seen using steady-state measurements and is independent of the repetition rate of the protocol.

We now test our ability to capture and store itinerant microwave states in the mechanical oscillator and to retrieve and measure those states after storage, as depicted in Fig. 3a. We simultaneously apply a transfer field with $\Gamma=2\pi \times 1.81\; \mathrm{kHz} > n_{\mathrm{env}} \gamma_\mathrm{m}$ and a coherent preparation pulse, such that the mechanical oscillator captures the preparation field. Although our preparation pulse has a square envelope, we capture a coherent temporal mode that is a growing exponential (time reverse of our retrieved state), chosen by the envelope of the transfer field (see Supplementary Information). After waiting a delay time $\tau_{\mathrm{del}}$ the mechanical state is transferred back into a microwave state and measured (Fig 3b). In order to demonstrate that the phase of the preparation field is preserved in the mechanical oscillator, we choose one of four different preparation fields, where the four choices have different phases. The data clearly show that the phase of the preparation is faithfully recovered after storage. Furthermore, the total variances in Fig.~3b are 4.5 times smaller than in Fig. 2b, signifying a lower entropy and thermal energy.

\begin{figure*}
  \centering
  \includegraphics[width= 5 in]{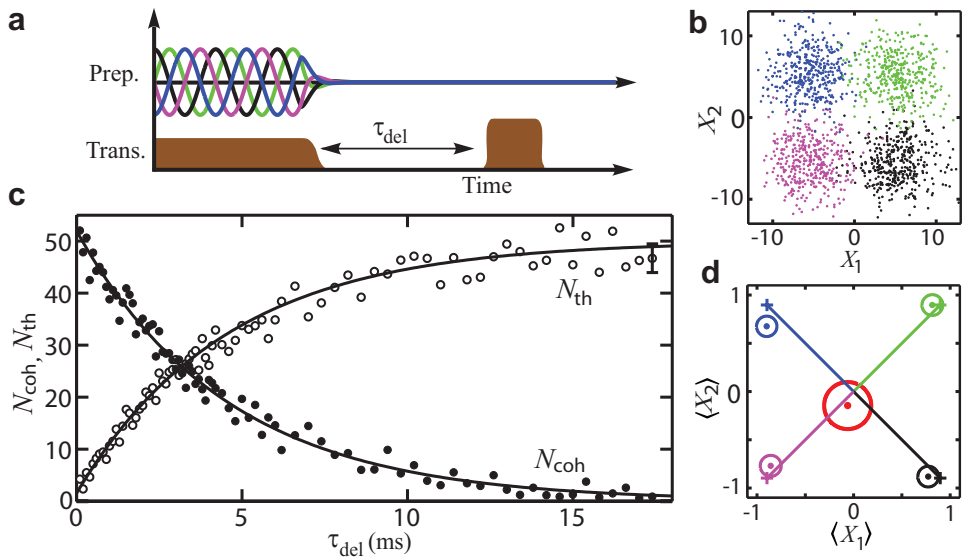}\\
  \caption{\textbf{A mechanical oscillator as a phase coherent memory. a,} The diagram depicts the protocol for demonstrating capture, storage, and retrieval of itinerant microwave fields. The upper trace shows the four possible coherent preparation fields as sinusoidal oscillations with one of four different phases, indicated by the four colours. The lower trace illustrates the strength of the transfer field. We repeat the protocol 1600 times, 400 times per phase. \textbf {b,} The state of the oscillator is inferred as in Fig. 2b with $\tau_{\mathrm{del}}=200\ \mu$s. Each measurement yields a single point in phase space that has been coloured to correspond to the phase of the preparation field. \textbf{c,} Occupancy of the oscillator as a function of $\tau_{\mathrm{del}}$ is decomposed into a thermal component (open circles) and a coherent component (closed circles). The solid lines are exponentials with rates independently determined from spectroscopic measurements, namely $\gamma_\mathrm{m} = 2\pi \times 35$ Hz. The error bar shown on the last point of the thermal component shows the expected one-standard deviation statistical uncertainty, $(N_\mathrm{th}+1/\eta)/\sqrt{400}$. \textbf{d,} To demonstrate coherent memory at the level of a single quantum, we repeat the protocol but use preparation fields with coherent amplitudes 6 times smaller than in \textbf {b}. The points show the coherent amplitude and phase of the retrieved state, estimated by averaging the quadrature values of 1000 measurements for each preparation phase. The circles enclosing the points indicate the one standard deviation statistical uncertainty in the measurement. For the outer points (coloured as in \textbf {b}), $\tau_{\mathrm{del}}=100\ \mu$s, where expected results are plotted as crosses in phase space. For the inner red point and circle, $\tau_{\mathrm{del}}=20$~ms, illustrating the eventual loss of phase information and associated increase in thermal noise.
  }\label{fig:fig3}
\end{figure*}

Having demonstrated the ability to prepare the mechanical oscillator in a low entropy state, we can observe its evolution back into a thermal state in equilibrium with its environment. In an ensemble of measurements we can distinguish the energy associated with the coherent and thermal components as $N_\mathrm{coh}=\langle X_1\rangle^2 + \langle X_2\rangle^2$ and $N_\mathrm{th}=\langle (X_1-\langle X_1 \rangle )^2 \rangle + \langle (X_2-\langle X_2 \rangle )^2 \rangle - 1/\eta$. Fig. 3c shows the evolution of $N_\mathrm{coh}$ and $N_\mathrm{th}$ as a function of $\tau_{\mathrm{del}}$. As expected, the coherent component exponentially decays while the entropy returns to the mechanical oscillator. We find excellent agreement between the data and predicted evolution with a rethermalization rate independently determined from steady-state measurements of the mechanical oscillator, $\gamma_\mathrm{m}=2\pi \times 35$ Hz. Thus, the rate at which the first thermal phonon enters the mechanical oscillator is well described by $n_{\mathrm{env}} \gamma_\mathrm{m} = 2\pi \times 1.75$ kHz, implying a storage time for a quantum memory of 90 $\mu$s.

Using our device as a quantum memory will require the ability to transfer states with about a single quantum of energy. To test our transfer process at the single quantum level we reduce the itinerant mode power by a factor of 36 from that of Fig. 3b and set $\tau_{\mathrm{del}}=100 \; \mathrm{\mu s}$. Based on our previous measurements of Fig. 3b, we would expect this protocol to prepare the mechanical oscillator in a state with $N_\mathrm{coh}=1.6$ quanta. The crosses in Fig. 3d show the expected coherent quadrature components $\langle X_1 \rangle$ and $\langle X_2 \rangle$, while the four outer points show the measured values of $\langle X_1 \rangle$ and $\langle X_2 \rangle$. We find that the state has been transferred into and out of the mechanical oscillator with an accuracy better than one quantum. The initial measurement of the oscillator at $t_\mathrm{del}=100\; \mu$s yields $N_\mathrm{th}= 2.14 \pm 0.14$ (see Supplementary Information). From the measured rethermalization rate $n_{\mathrm{env}} \gamma_\mathrm{m}$,  we can infer that the mechanics had one quantum of noise just after preparing it, i.e. $N_\mathrm{th}(\tau_\mathrm{del}=0 \; \mathrm{\mu s})\simeq 1$, consistent with the transfer field used to capture the preparation field.

\begin{figure}
  \centering
  \includegraphics[width= 3 in]{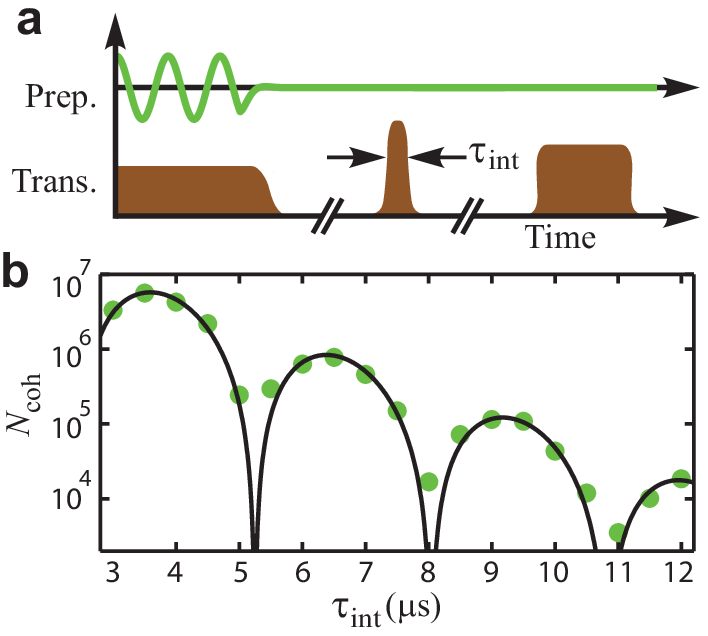}\\
  \caption{\textbf{State transfer between the resonator and oscillator. a,} The oscillator is first prepared in a large coherent state. An exchange transfer pulse with a large amplitude is then applied for a variable time $\tau_{\mathrm{int}}$, causing the states of the oscillator and resonator to swap with frequency $g_\mathrm{0} \sqrt{N_\mathrm{d}}/\pi$. Finally a lower amplitude measurement transfer field is applied to measure the state of the oscillator. \textbf{b,} The coherent component of the oscillator as a function of $\tau_{\mathrm{int}}$ (green points) clearly shows the expected swap oscillations. The solid line is a fit to the dynamics of the linearized interaction Hamiltonian (Eq.~\ref{eq:LinInteractHamil}).
  }\label{fig:fig4}
\end{figure}

While the minimum transfer rate to preserve quantum coherence between mechanics and itinerant microwave fields is $n_\mathrm{env} \gamma_\mathrm{m}$, the maximum transfer rate is $\kappa$. We illustrate this limit by increasing the transfer field strength such that $\sqrt{N_\mathrm{d}}> \kappa/2g_\mathrm{0}$, reaching a regime of state transfer between the mechanical oscillator and the resonator itself, rather than the itinerant fields that couple to the resonator. Specifically, we follow the protocol of Fig. 3a with the addition of a large transfer pulse (exchange) applied for variable time $\tau_{\mathrm{int}}$ between the preparation and measurement pulses (Fig. 4a). By waiting 100~$\mu$s after the exchange pulse to apply the measurement pulse, we measure only the portion of the state that was in the mechanical oscillator, as the resonator's energy decays much more rapidly. Figure 4b shows $N_\mathrm{coh}$ measured as a function of $\tau_{\mathrm{int}}$. Several exchanges between the two systems are visible, where the exchange rate is clearly faster than $\kappa$. The solid line is a fit using the equations of motion for a linearized model. We find good agreement for $\kappa=2\pi \times 220$ kHz and a coupling rate of $2g_\mathrm{0} \sqrt{N_\mathrm{d}}=1.7\;\kappa$. This measurement demonstrates in the time domain the bandwidth limit for transferring itinerant fields and that the ratio of transfer time to storage time is approximately 125.

Looking forward, the ability to store microwave states in a low loss mechanical oscillator provides an attractive tool in the growing field of quantum technology. While our measurements here have dealt with classical states with single quanta level energies, one can readily imagine incorporating a superconducting qubit to generate itinerant Fock states that could be stored in the mechanical oscillator \cite{Houck2007, Eichler2011}. In this case, the transfer field would be shaped to best capture the single itinerant photon \cite{Novikova2007}. Our measurements also represent a significant step in a more ambitious goal of transferring states between photons at microwave and optical frequencies \cite{Wang2012, Safavi-Naeini2011, Regal2011}. As mechanical oscillators continue to progress in the quantum regime, new applications will emerge that exploit this quantum behavior in the topics of fundamental physics, quantum information, and precise force sensing. Our measurements demonstrate a unique ability to control a mechanical oscillator, paving the way for many of these applications to become reality.

\begin{acknowledgments}
This work was supported primarily by the DARPA QuASAR programme, with additional support from the NSF Physics Frontier Center and NIST. We would like to thank Joseph Kerckhoff and Reed Andrews for fruitful discussions.

\end{acknowledgments}

\bibliography{StateTransfer}

\begin{thebibliography}{32}
\expandafter\ifx\csname natexlab\endcsname\relax\def\natexlab#1{#1}\fi
\expandafter\ifx\csname bibnamefont\endcsname\relax
  \def\bibnamefont#1{#1}\fi
\expandafter\ifx\csname bibfnamefont\endcsname\relax
  \def\bibfnamefont#1{#1}\fi
\expandafter\ifx\csname citenamefont\endcsname\relax
  \def\citenamefont#1{#1}\fi
\expandafter\ifx\csname url\endcsname\relax
  \def\url#1{\texttt{#1}}\fi
\expandafter\ifx\csname urlprefix\endcsname\relax\def\urlprefix{URL }\fi
\providecommand{\bibinfo}[2]{#2}
\providecommand{\eprint}[2][]{\url{#2}}

\bibitem[{\citenamefont{O'Connell et~al.}(2010)\citenamefont{O'Connell,
  Hofheinz, Ansmann, Bialczak, Lenander, Lucero, Neeley, Sank, Wang, Weides
  et~al.}}]{O'Connell2010}
\bibinfo{author}{\bibfnamefont{A.~D.} \bibnamefont{O'Connell}},
  \bibinfo{author}{\bibfnamefont{M.}~\bibnamefont{Hofheinz}},
  \bibinfo{author}{\bibfnamefont{M.}~\bibnamefont{Ansmann}},
  \bibinfo{author}{\bibfnamefont{R.~C.} \bibnamefont{Bialczak}},
  \bibinfo{author}{\bibfnamefont{M.}~\bibnamefont{Lenander}},
  \bibinfo{author}{\bibfnamefont{E.}~\bibnamefont{Lucero}},
  \bibinfo{author}{\bibfnamefont{M.}~\bibnamefont{Neeley}},
  \bibinfo{author}{\bibfnamefont{D.}~\bibnamefont{Sank}},
  \bibinfo{author}{\bibfnamefont{H.}~\bibnamefont{Wang}},
  \bibinfo{author}{\bibfnamefont{M.}~\bibnamefont{Weides}},
  \bibnamefont{et~al.}, \bibinfo{journal}{Nature}
  \textbf{\bibinfo{volume}{464}}, \bibinfo{pages}{697} (\bibinfo{year}{2010}).

\bibitem[{\citenamefont{Teufel et~al.}(2011{\natexlab{a}})\citenamefont{Teufel,
  Donner, Li, Harlow, Allman, Cicak, Sirois, Whittaker, Lehnert, and
  Simmonds}}]{Teufel2011b}
\bibinfo{author}{\bibfnamefont{J.~D.} \bibnamefont{Teufel}},
  \bibinfo{author}{\bibfnamefont{T.}~\bibnamefont{Donner}},
  \bibinfo{author}{\bibfnamefont{D.}~\bibnamefont{Li}},
  \bibinfo{author}{\bibfnamefont{J.~W.} \bibnamefont{Harlow}},
  \bibinfo{author}{\bibfnamefont{M.~S.} \bibnamefont{Allman}},
  \bibinfo{author}{\bibfnamefont{K.}~\bibnamefont{Cicak}},
  \bibinfo{author}{\bibfnamefont{A.~J.} \bibnamefont{Sirois}},
  \bibinfo{author}{\bibfnamefont{J.~D.} \bibnamefont{Whittaker}},
  \bibinfo{author}{\bibfnamefont{K.~W.} \bibnamefont{Lehnert}},
  \bibnamefont{and} \bibinfo{author}{\bibfnamefont{R.~W.}
  \bibnamefont{Simmonds}}, \bibinfo{journal}{Nature}
  \textbf{\bibinfo{volume}{475}}, \bibinfo{pages}{359–}
  (\bibinfo{year}{2011}{\natexlab{a}}).

\bibitem[{\citenamefont{Chan et~al.}(2011)\citenamefont{Chan, Mayer-Alegre,
  Safavi-Naeini, Hill, Krause, Groblacher, Aspelmeyer, and Painter}}]{Chan2011}
\bibinfo{author}{\bibfnamefont{J.}~\bibnamefont{Chan}},
  \bibinfo{author}{\bibfnamefont{T.~P.} \bibnamefont{Mayer-Alegre}},
  \bibinfo{author}{\bibfnamefont{A.~H.} \bibnamefont{Safavi-Naeini}},
  \bibinfo{author}{\bibfnamefont{J.~T.} \bibnamefont{Hill}},
  \bibinfo{author}{\bibfnamefont{A.}~\bibnamefont{Krause}},
  \bibinfo{author}{\bibfnamefont{S.}~\bibnamefont{Groblacher}},
  \bibinfo{author}{\bibfnamefont{M.}~\bibnamefont{Aspelmeyer}},
  \bibnamefont{and} \bibinfo{author}{\bibfnamefont{O.}~\bibnamefont{Painter}},
  \bibinfo{journal}{Nature} \textbf{\bibinfo{volume}{478}}, \bibinfo{pages}{89}
  (\bibinfo{year}{2011}).

\bibitem[{\citenamefont{Hertzberg et~al.}(2010)\citenamefont{Hertzberg,
  Rocheleau, Ndukum, Savva, Clerk, and Schwab}}]{Hertzberg2010}
\bibinfo{author}{\bibfnamefont{J.~B.} \bibnamefont{Hertzberg}},
  \bibinfo{author}{\bibfnamefont{T.}~\bibnamefont{Rocheleau}},
  \bibinfo{author}{\bibfnamefont{T.}~\bibnamefont{Ndukum}},
  \bibinfo{author}{\bibfnamefont{M.}~\bibnamefont{Savva}},
  \bibinfo{author}{\bibfnamefont{A.~A.} \bibnamefont{Clerk}}, \bibnamefont{and}
  \bibinfo{author}{\bibfnamefont{K.~C.} \bibnamefont{Schwab}},
  \bibinfo{journal}{Nature Physics} \textbf{\bibinfo{volume}{6}},
  \bibinfo{pages}{213} (\bibinfo{year}{2010}).

\bibitem[{\citenamefont{Sillanpaa et~al.}(2009)\citenamefont{Sillanpaa, Sarkar,
  Sulkko, Muhonen, and Hakonen}}]{Sillanpaa2009}
\bibinfo{author}{\bibfnamefont{M.~A.} \bibnamefont{Sillanpaa}},
  \bibinfo{author}{\bibfnamefont{J.}~\bibnamefont{Sarkar}},
  \bibinfo{author}{\bibfnamefont{J.}~\bibnamefont{Sulkko}},
  \bibinfo{author}{\bibfnamefont{J.}~\bibnamefont{Muhonen}}, \bibnamefont{and}
  \bibinfo{author}{\bibfnamefont{P.~J.} \bibnamefont{Hakonen}},
  \bibinfo{journal}{Appl. Phys. Lett.} \textbf{\bibinfo{volume}{95}},
  \bibinfo{pages}{011909} (\bibinfo{year}{2009}).

\bibitem[{\citenamefont{LaHaye et~al.}(2009)\citenamefont{LaHaye, Suh,
  Echternach, Schwab, and Roukes}}]{LaHaye2009}
\bibinfo{author}{\bibfnamefont{M.~D.} \bibnamefont{LaHaye}},
  \bibinfo{author}{\bibfnamefont{J.}~\bibnamefont{Suh}},
  \bibinfo{author}{\bibfnamefont{P.~M.} \bibnamefont{Echternach}},
  \bibinfo{author}{\bibfnamefont{K.~C.} \bibnamefont{Schwab}},
  \bibnamefont{and} \bibinfo{author}{\bibfnamefont{M.~L.}
  \bibnamefont{Roukes}}, \bibinfo{journal}{Nature}
  \textbf{\bibinfo{volume}{459}}, \bibinfo{pages}{960} (\bibinfo{year}{2009}).

\bibitem[{\citenamefont{Regal et~al.}(2008)\citenamefont{Regal, Teufel, and
  Lehnert}}]{REGAL2008}
\bibinfo{author}{\bibfnamefont{C.~A.} \bibnamefont{Regal}},
  \bibinfo{author}{\bibfnamefont{J.~D.} \bibnamefont{Teufel}},
  \bibnamefont{and} \bibinfo{author}{\bibfnamefont{K.~W.}
  \bibnamefont{Lehnert}}, \bibinfo{journal}{Nature Physics}
  \textbf{\bibinfo{volume}{4}}, \bibinfo{pages}{555} (\bibinfo{year}{2008}).

\bibitem[{\citenamefont{Wang and Clerk}(2012)}]{Wang2012}
\bibinfo{author}{\bibfnamefont{Y.-D.} \bibnamefont{Wang}} \bibnamefont{and}
  \bibinfo{author}{\bibfnamefont{A.~A.} \bibnamefont{Clerk}},
  \bibinfo{journal}{Phys. Rev. Lett.} \textbf{\bibinfo{volume}{108}},
  \bibinfo{pages}{153603} (\bibinfo{year}{2012}).

\bibitem[{\citenamefont{Regal and Lehnert}(2011)}]{Regal2011}
\bibinfo{author}{\bibfnamefont{C.~A.} \bibnamefont{Regal}} \bibnamefont{and}
  \bibinfo{author}{\bibfnamefont{K.~W.} \bibnamefont{Lehnert}},
  \bibinfo{journal}{Journal of Physics: Conference Series}
  \textbf{\bibinfo{volume}{264}}, \bibinfo{pages}{1799} (\bibinfo{year}{2011}).

\bibitem[{\citenamefont{Houck et~al.}(2007)\citenamefont{Houck, Schuster,
  Gambetta, Schreier, Johnson, Chow, Frunzio, Majer, Devoret, Girvin
  et~al.}}]{Houck2007}
\bibinfo{author}{\bibfnamefont{A.~A.} \bibnamefont{Houck}},
  \bibinfo{author}{\bibfnamefont{D.~I.} \bibnamefont{Schuster}},
  \bibinfo{author}{\bibfnamefont{J.~M.} \bibnamefont{Gambetta}},
  \bibinfo{author}{\bibfnamefont{J.~A.} \bibnamefont{Schreier}},
  \bibinfo{author}{\bibfnamefont{B.~R.} \bibnamefont{Johnson}},
  \bibinfo{author}{\bibfnamefont{J.~M.} \bibnamefont{Chow}},
  \bibinfo{author}{\bibfnamefont{L.}~\bibnamefont{Frunzio}},
  \bibinfo{author}{\bibfnamefont{J.}~\bibnamefont{Majer}},
  \bibinfo{author}{\bibfnamefont{M.~H.} \bibnamefont{Devoret}},
  \bibinfo{author}{\bibfnamefont{S.~M.} \bibnamefont{Girvin}},
  \bibnamefont{et~al.}, \bibinfo{journal}{Nature}
  \textbf{\bibinfo{volume}{449}}, \bibinfo{pages}{328} (\bibinfo{year}{2007}).

\bibitem[{\citenamefont{Mallet et~al.}(2011)\citenamefont{Mallet,
  Castellanos-Beltran, Ku, Glancy, Knill, Irwin, Hilton, Vale, and
  Lehnert}}]{Mallet2011}
\bibinfo{author}{\bibfnamefont{F.}~\bibnamefont{Mallet}},
  \bibinfo{author}{\bibfnamefont{M.~A.} \bibnamefont{Castellanos-Beltran}},
  \bibinfo{author}{\bibfnamefont{H.~S.} \bibnamefont{Ku}},
  \bibinfo{author}{\bibfnamefont{S.}~\bibnamefont{Glancy}},
  \bibinfo{author}{\bibfnamefont{E.}~\bibnamefont{Knill}},
  \bibinfo{author}{\bibfnamefont{K.~D.} \bibnamefont{Irwin}},
  \bibinfo{author}{\bibfnamefont{G.~C.} \bibnamefont{Hilton}},
  \bibinfo{author}{\bibfnamefont{L.~R.} \bibnamefont{Vale}}, \bibnamefont{and}
  \bibinfo{author}{\bibfnamefont{K.~W.} \bibnamefont{Lehnert}},
  \bibinfo{journal}{Phys. Rev. Lett.} \textbf{\bibinfo{volume}{106}},
  \bibinfo{pages}{220502} (\bibinfo{year}{2011}).

\bibitem[{\citenamefont{Eichler et~al.}(2011)\citenamefont{Eichler, Bozyigit,
  Lang, Steffen, Fink, and Wallraff}}]{Eichler2011}
\bibinfo{author}{\bibfnamefont{C.}~\bibnamefont{Eichler}},
  \bibinfo{author}{\bibfnamefont{D.}~\bibnamefont{Bozyigit}},
  \bibinfo{author}{\bibfnamefont{C.}~\bibnamefont{Lang}},
  \bibinfo{author}{\bibfnamefont{L.}~\bibnamefont{Steffen}},
  \bibinfo{author}{\bibfnamefont{J.}~\bibnamefont{Fink}}, \bibnamefont{and}
  \bibinfo{author}{\bibfnamefont{A.}~\bibnamefont{Wallraff}},
  \bibinfo{journal}{Phys. Rev. Lett.} \textbf{\bibinfo{volume}{106}},
  \bibinfo{pages}{220503} (\bibinfo{year}{2011}).

\bibitem[{\citenamefont{Cicak et~al.}(2010)\citenamefont{Cicak, Li, Strong,
  Allman, Altomare, Sirois, Whittaker, Teufel, and Simmonds}}]{Cicak2010}
\bibinfo{author}{\bibfnamefont{K.}~\bibnamefont{Cicak}},
  \bibinfo{author}{\bibfnamefont{D.}~\bibnamefont{Li}},
  \bibinfo{author}{\bibfnamefont{J.~A.} \bibnamefont{Strong}},
  \bibinfo{author}{\bibfnamefont{M.~S.} \bibnamefont{Allman}},
  \bibinfo{author}{\bibfnamefont{F.}~\bibnamefont{Altomare}},
  \bibinfo{author}{\bibfnamefont{A.~J.} \bibnamefont{Sirois}},
  \bibinfo{author}{\bibfnamefont{J.~D.} \bibnamefont{Whittaker}},
  \bibinfo{author}{\bibfnamefont{J.~D.} \bibnamefont{Teufel}},
  \bibnamefont{and} \bibinfo{author}{\bibfnamefont{R.~W.}
  \bibnamefont{Simmonds}}, \bibinfo{journal}{Journal of Applied Physics}
  \textbf{\bibinfo{volume}{96}}, \bibinfo{pages}{093502}
  (\bibinfo{year}{2010}).

\bibitem[{\citenamefont{Wang et~al.}(2011)\citenamefont{Wang, Vinjanampathy,
  Strauch, and Jacobs}}]{JacobsK2011}
\bibinfo{author}{\bibfnamefont{X.}~\bibnamefont{Wang}},
  \bibinfo{author}{\bibfnamefont{S.}~\bibnamefont{Vinjanampathy}},
  \bibinfo{author}{\bibfnamefont{F.~W.} \bibnamefont{Strauch}},
  \bibnamefont{and} \bibinfo{author}{\bibfnamefont{K.}~\bibnamefont{Jacobs}},
  \bibinfo{journal}{Phys. Rev. Lett.} \textbf{\bibinfo{volume}{107}},
  \bibinfo{pages}{177204} (\bibinfo{year}{2011}).

\bibitem[{\citenamefont{Hofer et~al.}(2011)\citenamefont{Hofer, Wieczorek,
  Aspelmeyer, and Hammerer}}]{Hofer2011}
\bibinfo{author}{\bibfnamefont{S.~G.} \bibnamefont{Hofer}},
  \bibinfo{author}{\bibfnamefont{W.}~\bibnamefont{Wieczorek}},
  \bibinfo{author}{\bibfnamefont{M.}~\bibnamefont{Aspelmeyer}},
  \bibnamefont{and} \bibinfo{author}{\bibfnamefont{K.}~\bibnamefont{Hammerer}},
  \bibinfo{journal}{Phys. Rev. A} \textbf{\bibinfo{volume}{84}},
  \bibinfo{pages}{052327} (\bibinfo{year}{2011}).

\bibitem[{\citenamefont{Romero-Isart et~al.}(2011)\citenamefont{Romero-Isart,
  Pflanzer, Blaser, Kaltenbaek, Kiesel, Aspelmeyer, and Cirac}}]{Romero2011}
\bibinfo{author}{\bibfnamefont{O.}~\bibnamefont{Romero-Isart}},
  \bibinfo{author}{\bibfnamefont{A.~C.} \bibnamefont{Pflanzer}},
  \bibinfo{author}{\bibfnamefont{F.}~\bibnamefont{Blaser}},
  \bibinfo{author}{\bibfnamefont{R.}~\bibnamefont{Kaltenbaek}},
  \bibinfo{author}{\bibfnamefont{N.}~\bibnamefont{Kiesel}},
  \bibinfo{author}{\bibfnamefont{M.}~\bibnamefont{Aspelmeyer}},
  \bibnamefont{and} \bibinfo{author}{\bibfnamefont{J.~I.} \bibnamefont{Cirac}},
  \bibinfo{journal}{Phys. Rev. Lett.} \textbf{\bibinfo{volume}{107}},
  \bibinfo{pages}{020405} (\bibinfo{year}{2011}).

\bibitem[{\citenamefont{Machnes et~al.}(2012)\citenamefont{Machnes, Cerrillo,
  Aspelmeyer, Wieczorek, Plenio, and Retzker}}]{Machnes2012}
\bibinfo{author}{\bibfnamefont{S.}~\bibnamefont{Machnes}},
  \bibinfo{author}{\bibfnamefont{J.}~\bibnamefont{Cerrillo}},
  \bibinfo{author}{\bibfnamefont{M.}~\bibnamefont{Aspelmeyer}},
  \bibinfo{author}{\bibfnamefont{W.}~\bibnamefont{Wieczorek}},
  \bibinfo{author}{\bibfnamefont{M.~B.} \bibnamefont{Plenio}},
  \bibnamefont{and} \bibinfo{author}{\bibfnamefont{A.}~\bibnamefont{Retzker}},
  \bibinfo{journal}{Phys. Rev. Lett.} \textbf{\bibinfo{volume}{108}},
  \bibinfo{pages}{153601} (\bibinfo{year}{2012}).

\bibitem[{\citenamefont{Vanner et~al.}(2011)\citenamefont{Vanner, Pikovski,
  Cole, Kim, Brukner, Hammerer, Milburn, and Aspelmeyer}}]{Vanner2011}
\bibinfo{author}{\bibfnamefont{M.~R.} \bibnamefont{Vanner}},
  \bibinfo{author}{\bibfnamefont{I.}~\bibnamefont{Pikovski}},
  \bibinfo{author}{\bibfnamefont{G.~D.} \bibnamefont{Cole}},
  \bibinfo{author}{\bibfnamefont{M.~S.} \bibnamefont{Kim}},
  \bibinfo{author}{\bibfnamefont{c.}~\bibnamefont{Brukner}},
  \bibinfo{author}{\bibfnamefont{K.}~\bibnamefont{Hammerer}},
  \bibinfo{author}{\bibfnamefont{G.~J.} \bibnamefont{Milburn}},
  \bibnamefont{and}
  \bibinfo{author}{\bibfnamefont{M.}~\bibnamefont{Aspelmeyer}},
  \bibinfo{journal}{PNAS USA} \textbf{\bibinfo{volume}{108}},
  \bibinfo{pages}{16182} (\bibinfo{year}{2011}).

\bibitem[{\citenamefont{Groblacher et~al.}(2009)\citenamefont{Groblacher,
  Hammerer, Vanner, and Aspelmeyer}}]{Groblacher2009a}
\bibinfo{author}{\bibfnamefont{S.}~\bibnamefont{Groblacher}},
  \bibinfo{author}{\bibfnamefont{K.}~\bibnamefont{Hammerer}},
  \bibinfo{author}{\bibfnamefont{M.~R.} \bibnamefont{Vanner}},
  \bibnamefont{and}
  \bibinfo{author}{\bibfnamefont{M.}~\bibnamefont{Aspelmeyer}},
  \bibinfo{journal}{Nature} \textbf{\bibinfo{volume}{460}},
  \bibinfo{pages}{724} (\bibinfo{year}{2009}).

\bibitem[{\citenamefont{Gao et~al.}(2008)\citenamefont{Gao, Daal, Vayonakis,
  Kumar, Zmuidzinas, Sadoulet, Mazin, Day, and Leduc}}]{Gao2008a}
\bibinfo{author}{\bibfnamefont{J.~S.} \bibnamefont{Gao}},
  \bibinfo{author}{\bibfnamefont{M.}~\bibnamefont{Daal}},
  \bibinfo{author}{\bibfnamefont{A.}~\bibnamefont{Vayonakis}},
  \bibinfo{author}{\bibfnamefont{S.}~\bibnamefont{Kumar}},
  \bibinfo{author}{\bibfnamefont{J.}~\bibnamefont{Zmuidzinas}},
  \bibinfo{author}{\bibfnamefont{B.}~\bibnamefont{Sadoulet}},
  \bibinfo{author}{\bibfnamefont{B.~A.} \bibnamefont{Mazin}},
  \bibinfo{author}{\bibfnamefont{P.~K.} \bibnamefont{Day}}, \bibnamefont{and}
  \bibinfo{author}{\bibfnamefont{H.~G.} \bibnamefont{Leduc}},
  \bibinfo{journal}{Applied Physics Letters} \textbf{\bibinfo{volume}{92}}
  (\bibinfo{year}{2008}).

\bibitem[{\citenamefont{Fiore et~al.}(2011)\citenamefont{Fiore, Yang, Kuzyk,
  Barbour, Tian, and Wang}}]{Fiore2011}
\bibinfo{author}{\bibfnamefont{V.}~\bibnamefont{Fiore}},
  \bibinfo{author}{\bibfnamefont{Y.}~\bibnamefont{Yang}},
  \bibinfo{author}{\bibfnamefont{M.~C.} \bibnamefont{Kuzyk}},
  \bibinfo{author}{\bibfnamefont{R.}~\bibnamefont{Barbour}},
  \bibinfo{author}{\bibfnamefont{L.}~\bibnamefont{Tian}}, \bibnamefont{and}
  \bibinfo{author}{\bibfnamefont{H.}~\bibnamefont{Wang}},
  \bibinfo{journal}{Phys. Rev. Lett.} \textbf{\bibinfo{volume}{107}},
  \bibinfo{pages}{133601} (\bibinfo{year}{2011}).

\bibitem[{\citenamefont{Chang et~al.}(2011)\citenamefont{Chang, Safavi-Naeini,
  Hafezi, and Painter}}]{Chang2011}
\bibinfo{author}{\bibfnamefont{D.~E.} \bibnamefont{Chang}},
  \bibinfo{author}{\bibfnamefont{A.~H.} \bibnamefont{Safavi-Naeini}},
  \bibinfo{author}{\bibfnamefont{M.}~\bibnamefont{Hafezi}}, \bibnamefont{and}
  \bibinfo{author}{\bibfnamefont{O.}~\bibnamefont{Painter}},
  \bibinfo{journal}{New Journal of Physics} \textbf{\bibinfo{volume}{13}},
  \bibinfo{pages}{023003} (\bibinfo{year}{2011}).

\bibitem[{\citenamefont{Verhagen et~al.}(2012)\citenamefont{Verhagen,
  Deleglise, Weis, Schliesser, and Kippenberg}}]{Verhagen2012}
\bibinfo{author}{\bibfnamefont{E.}~\bibnamefont{Verhagen}},
  \bibinfo{author}{\bibfnamefont{S.}~\bibnamefont{Deleglise}},
  \bibinfo{author}{\bibfnamefont{S.}~\bibnamefont{Weis}},
  \bibinfo{author}{\bibfnamefont{A.}~\bibnamefont{Schliesser}},
  \bibnamefont{and} \bibinfo{author}{\bibfnamefont{T.~J.}
  \bibnamefont{Kippenberg}}, \bibinfo{journal}{Nature}
  \textbf{\bibinfo{volume}{482}}, \bibinfo{pages}{63} (\bibinfo{year}{2012}).

\bibitem[{\citenamefont{Teufel et~al.}(2011{\natexlab{b}})\citenamefont{Teufel,
  Li, Allman, Cicak, Sirois, Whittaker, and Simmonds}}]{Teufel2011}
\bibinfo{author}{\bibfnamefont{J.~D.} \bibnamefont{Teufel}},
  \bibinfo{author}{\bibfnamefont{D.}~\bibnamefont{Li}},
  \bibinfo{author}{\bibfnamefont{M.~S.} \bibnamefont{Allman}},
  \bibinfo{author}{\bibfnamefont{K.}~\bibnamefont{Cicak}},
  \bibinfo{author}{\bibfnamefont{A.~J.} \bibnamefont{Sirois}},
  \bibinfo{author}{\bibfnamefont{J.~D.} \bibnamefont{Whittaker}},
  \bibnamefont{and} \bibinfo{author}{\bibfnamefont{R.~W.}
  \bibnamefont{Simmonds}}, \bibinfo{journal}{Nature}
  \textbf{\bibinfo{volume}{471}}, \bibinfo{pages}{204}
  (\bibinfo{year}{2011}{\natexlab{b}}).

\bibitem[{\citenamefont{Teufel et~al.}(2008)\citenamefont{Teufel, Harlow,
  Regal, and Lehnert}}]{Teufel2008a}
\bibinfo{author}{\bibfnamefont{J.~D.} \bibnamefont{Teufel}},
  \bibinfo{author}{\bibfnamefont{J.~W.} \bibnamefont{Harlow}},
  \bibinfo{author}{\bibfnamefont{C.~A.} \bibnamefont{Regal}}, \bibnamefont{and}
  \bibinfo{author}{\bibfnamefont{K.~W.} \bibnamefont{Lehnert}},
  \bibinfo{journal}{Physical Review Letters} \textbf{\bibinfo{volume}{101}},
  \bibinfo{pages}{197203} (\bibinfo{year}{2008}).

\bibitem[{\citenamefont{Rocheleau et~al.}(2010)\citenamefont{Rocheleau, Ndukum,
  Macklin, Hertzberg, Clerk, and Schwab}}]{Rocheleau2010}
\bibinfo{author}{\bibfnamefont{T.}~\bibnamefont{Rocheleau}},
  \bibinfo{author}{\bibfnamefont{T.}~\bibnamefont{Ndukum}},
  \bibinfo{author}{\bibfnamefont{C.}~\bibnamefont{Macklin}},
  \bibinfo{author}{\bibfnamefont{J.~B.} \bibnamefont{Hertzberg}},
  \bibinfo{author}{\bibfnamefont{A.~A.} \bibnamefont{Clerk}}, \bibnamefont{and}
  \bibinfo{author}{\bibfnamefont{K.~C.} \bibnamefont{Schwab}},
  \bibinfo{journal}{Nature} \textbf{\bibinfo{volume}{463}}, \bibinfo{pages}{72}
  (\bibinfo{year}{2010}).

\bibitem[{\citenamefont{Marquardt et~al.}(2007)\citenamefont{Marquardt, Chen,
  Clerk, and Girvin}}]{Marquardt2007}
\bibinfo{author}{\bibfnamefont{F.}~\bibnamefont{Marquardt}},
  \bibinfo{author}{\bibfnamefont{J.~P.} \bibnamefont{Chen}},
  \bibinfo{author}{\bibfnamefont{A.~A.} \bibnamefont{Clerk}}, \bibnamefont{and}
  \bibinfo{author}{\bibfnamefont{S.~M.} \bibnamefont{Girvin}},
  \bibinfo{journal}{Physical Review Letters} \textbf{\bibinfo{volume}{99}},
  \bibinfo{pages}{093902} (\bibinfo{year}{2007}).

\bibitem[{\citenamefont{Wilson-Rae et~al.}(2007)\citenamefont{Wilson-Rae,
  Nooshi, Zwerger, and Kippenberg}}]{Wilson-Rae2007}
\bibinfo{author}{\bibfnamefont{I.}~\bibnamefont{Wilson-Rae}},
  \bibinfo{author}{\bibfnamefont{N.}~\bibnamefont{Nooshi}},
  \bibinfo{author}{\bibfnamefont{W.}~\bibnamefont{Zwerger}}, \bibnamefont{and}
  \bibinfo{author}{\bibfnamefont{T.~J.} \bibnamefont{Kippenberg}},
  \bibinfo{journal}{Phys. Rev. Lett.} \textbf{\bibinfo{volume}{99}},
  \bibinfo{pages}{093901} (\bibinfo{year}{2007}).

\bibitem[{\citenamefont{Safavi-Naeini et~al.}(2011)\citenamefont{Safavi-Naeini,
  Mayer~Alegre, Chan, Eichenfield, Winger, Lin, Hill, Chang, and
  Painter}}]{Safavi2011}
\bibinfo{author}{\bibfnamefont{A.~H.} \bibnamefont{Safavi-Naeini}},
  \bibinfo{author}{\bibfnamefont{T.~P.} \bibnamefont{Mayer~Alegre}},
  \bibinfo{author}{\bibfnamefont{J.}~\bibnamefont{Chan}},
  \bibinfo{author}{\bibfnamefont{M.}~\bibnamefont{Eichenfield}},
  \bibinfo{author}{\bibfnamefont{M.}~\bibnamefont{Winger}},
  \bibinfo{author}{\bibfnamefont{Q.}~\bibnamefont{Lin}},
  \bibinfo{author}{\bibfnamefont{J.~T.} \bibnamefont{Hill}},
  \bibinfo{author}{\bibfnamefont{D.~E.} \bibnamefont{Chang}}, \bibnamefont{and}
  \bibinfo{author}{\bibfnamefont{O.}~\bibnamefont{Painter}},
  \bibinfo{journal}{Nature} \textbf{\bibinfo{volume}{472}}, \bibinfo{pages}{69}
  (\bibinfo{year}{2011}).

\bibitem[{\citenamefont{Novikova et~al.}(2007)\citenamefont{Novikova, Gorshkov,
  Phillips, Sorensen, Lukin, and Walsworth}}]{Novikova2007}
\bibinfo{author}{\bibfnamefont{I.}~\bibnamefont{Novikova}},
  \bibinfo{author}{\bibfnamefont{A.~V.} \bibnamefont{Gorshkov}},
  \bibinfo{author}{\bibfnamefont{D.~F.} \bibnamefont{Phillips}},
  \bibinfo{author}{\bibfnamefont{A.~S.} \bibnamefont{Sorensen}},
  \bibinfo{author}{\bibfnamefont{M.~D.} \bibnamefont{Lukin}}, \bibnamefont{and}
  \bibinfo{author}{\bibfnamefont{R.~L.} \bibnamefont{Walsworth}},
  \bibinfo{journal}{Phys. Rev. Lett.} \textbf{\bibinfo{volume}{98}},
  \bibinfo{pages}{243602} (\bibinfo{year}{2007}).

\bibitem[{\citenamefont{Safavi-Naeini and Painter}(2011)}]{Safavi-Naeini2011}
\bibinfo{author}{\bibfnamefont{A.~H.} \bibnamefont{Safavi-Naeini}}
  \bibnamefont{and} \bibinfo{author}{\bibfnamefont{O.}~\bibnamefont{Painter}},
  \bibinfo{journal}{New J. Phys.} \textbf{\bibinfo{volume}{13}},
  \bibinfo{pages}{013017} (\bibinfo{year}{2011}).

\bibitem[{\citenamefont{Castellanos-Beltran
  et~al.}(2008)\citenamefont{Castellanos-Beltran, Irwin, Hilton, Vale, and
  Lehnert}}]{Castellanos-Beltran2008}
\bibinfo{author}{\bibfnamefont{M.~A.} \bibnamefont{Castellanos-Beltran}},
  \bibinfo{author}{\bibfnamefont{K.~D.} \bibnamefont{Irwin}},
  \bibinfo{author}{\bibfnamefont{G.~C.} \bibnamefont{Hilton}},
  \bibinfo{author}{\bibfnamefont{L.~R.} \bibnamefont{Vale}}, \bibnamefont{and}
  \bibinfo{author}{\bibfnamefont{K.~W.} \bibnamefont{Lehnert}},
  \bibinfo{journal}{Nature Phys.} \textbf{\bibinfo{volume}{4}},
  \bibinfo{pages}{929} (\bibinfo{year}{2008}).

\end{thebibliography}

\newpage
\begin{appendix}

\textbf{Supplementary Information}

\section{\textbf{Extracting the quadrature amplitudes and measurement efficiency}}

From each individual measurement we obtain a best estimate of the quadrature amplitudes $X_1$ and $X_2$ of the state of the mechanical oscillator. As shown in Fig. 2a, the output voltage $V_{\mathrm{out}}$ during the transfer pulse is an exponentially decaying oscillation obtained by amplifying and mixing down the itinerant microwave field.  The frequency component at $\omega_\mathrm{d}+\Omega_\mathrm{m}$ in the microwave field is mapped in the mixed down signal to $\omega_{\mathrm{out}} \approx 2\pi \times 1$~MHz. We digitally sample $V_{\mathrm{out}}$ at 10 MHz, yielding samples at discrete times $\{t_i\}$, which enables us to optimally filter the signal based on our knowledge of the transfer pulse and mechanical oscillator. The decay rate of $V_{\mathrm{out}}$ is controlled by the amplitude of the transfer pulse. We therefore determine our uncalibrated quadrature amplitudes by projecting the data onto the appropriate damped oscillatory functions as
\begin{equation}\label{eq:XX1}
X_1=C\sum_{i} e^{-\Gamma t_i/2} \cos((\omega_{\mathrm{out}}+\delta) t_i) V_{\mathrm{out}}(t_i) \; ,
\end{equation}
and
\begin{equation}\label{eq:XX2}
X_2=C\sum_{i} e^{-\Gamma t_i/2} \sin((\omega_{\mathrm{out}}+\delta) t_i) V_{\mathrm{out}}(t_i) \;,
\end{equation}
where $\delta$ is a frequency offset parameter and $C$ is a scaling parameter to convert our measurements into units of mechanical quanta. For the optimum estimate of $X_1$ and $X_2$, $\delta$ is set to 0.

While each individual estimate of $X_1$ and $X_2$  necessarily includes a component from both measurement noise and thermal noise, we can separately determine the variance of these two sources. The variance of the measurement noise can be found by projecting the data onto functions orthogonal to the template functions $e^{-\Gamma t/2} \cos(\omega_{\mathrm{out}} t)$ and $e^{-\Gamma t/2} \sin(\omega_{\mathrm{out}} t)$.  In practice we determine the uncalibrated measurement noise from the variance of $X_1$ and $X_2$ when Eqs.~\ref{eq:XX1} and \ref{eq:XX2} are evaluated with $|\delta| \gg \Gamma$, sufficiently large to be orthogonal to the template functions (Fig. 5). As the measurement noise and thermal noise are uncorrelated, we can subtract the variance of the measurement noise $1/\eta$ from the total variance of the ensemble measurements to obtain the variance associated with the thermal noise. As described in the main text, we scale $X_1$, $X_2$, and $1/\eta$ into calibrated units of quanta by preparing the mechanical oscillator in a state of known temperature (see Fig 2c).

In Fig. 5 we plot the total variance $N_\mathrm{th}+1/\eta =\langle (X_1-\langle X_1 \rangle )^2 \rangle + \langle (X_2-\langle X_2 \rangle )^2 \rangle$ versus $\delta$ for the 25~mK thermal state (red points) and a low entropy state (green points) also shown in Fig. 3d. The solid line is a Lorentzian fit to extract $1/\eta$ and $N_{\mathrm{th}}$. Here we find $1/\eta=0.11$ and $N_{\mathrm{th}}=2.22$ for the low entropy coherent state. Similar measurements on the other low entropy states shown in Fig. 3d give an average thermal component of $N_{\mathrm{th}}=2.14 \pm 0.14$.

\begin{figure}
  \centering
  \includegraphics[width= 3 in]{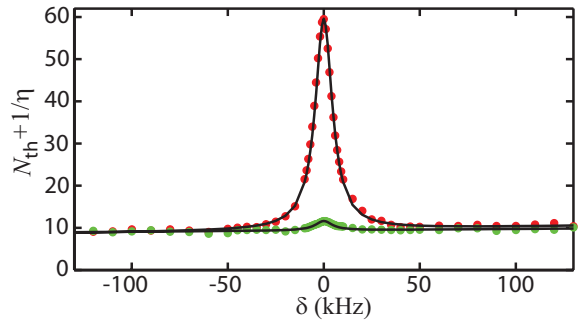}\\
  \caption{\textbf{Measurement efficiency.} The red points correspond to the variance of a 25 mK thermal state, while the green points correspond to a low entropy state, plotted as a function of $\delta$. The coherent components from the same measurements are depicted as the green and red points in Fig. 3d. The line is a Lorentzian fit used to extract $N_{\mathrm{th}}$ and $\eta$ from the peak and offset, respectively. The width of the lorentzians $\Gamma$ is set by the measurement strength $N_d$. Here we find an efficiency of $\eta=0.11$. The background slope results from a slight frequency dependence in the amplifier chain.
  }\label{fig:fig5}
\end{figure}

\section{\textbf{Predicted measurement efficiency}}

We infer the oscillator energy from simultaneous measurements of $X_1$ and $X_2$, and thus expect quantum noise to contribute in two different ways. First, neither $X_1$ nor $X_2$ commute with the Hamiltonian of the oscillator, accounting for the appearance of 1/2 quantum of noise, usually associated with oscillator zero-point motion. In addition, $X_1$ and $X_2$ do not commute with each other; the associated Heisenberg uncertainty principle contributes an additional 1/2 quantum of measurement noise. Therefore a quantum limited measurement of a thermal state will have variance $n_{\mathrm{th}}+1$. By writing the variance of our measurement as $n_{\mathrm{th}}+1/\eta$, $\eta$ can be regarded as the measurement's quantum efficiency. We specify the measurement noise as a quantum measurement efficiency $\eta$ to highlight the combined effects of loss and amplifier noise with respect to an ideal quantum limited experiment.

We can separately consider the factors contributing to a reduced efficiency and multiply them together to get a best estimate of $\eta$. These factors can be grouped in terms of those that affect the state transfer efficiency $\eta_{\mathrm{st}}$ and those that affect our inference of the state transferred. As $\kappa_{{\mathrm{int}}}$ is comparable to $\kappa_{\mathrm{ext}}$, 65 percent of the oscillator energy is transferred to itinerant photons, as opposed to being dissipated in the resonator. Also, in the current configuration we only measure the waves propagating in one direction, yet the resonator mode couples to modes propagating in two directions. These two factors combine to give an estimated $\eta_{\mathrm{st}} = 0.33$. Our inference of the transferred state is affected by loss in the microwave components and a non-quantum limited amplifier. The microwave components that carry the signal from the resonator to the Josephson Parametric Amplifier (JPA) transmit 55 percent of the energy and the added noise of the JPA \cite{Castellanos-Beltran2008} can be recast as a quantum efficiency of 70 percent \cite{Mallet2011}. When these factors are combined with $\eta_{\mathrm{st}}$ we expect a total measurement efficiency $\eta=0.13$.

In future experiments we should be able to achieve a significantly higher efficiency. We plan to couple the resonator to only one direction, improve the ratio of $\kappa/\kappa_{\mathrm{int}}$, and use a JPA closer to being quantum limited.  These factors should allow $\eta_{\mathrm{st}}$ to approach 1 and improve $\eta$ to approximately 0.5.

\section{\textbf{Temporal Mode Capture and Transfer Efficiency}}

\begin{figure}
  \centering
  \includegraphics[width= 3 in]{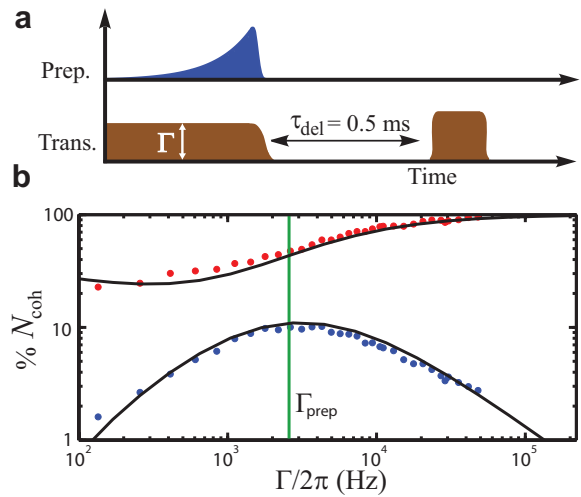}\\
  \caption{\textbf{Temporal Mode Capture. a,} The diagram depicts the protocol for capture, storage, and retrieval for an itinerant microwave field that ideally matches a constant transfer field. The upper trace shows the power of the preparation field, a rising exponential at rate $\Gamma_\mathrm{prep}$. \textbf{b}, The coherent microwave energy retrieved from the mechanical oscillator (blue) is plotted as a function of the transfer field strength $\Gamma$. The y-axis has been normalized to the total energy in the preparation field. The red points show the fraction of the preparation field that passes through the transmission line directly without being captured by the oscillator. The solid lines are the expected results with all parameters determined in previous measurements.
}\label{fig:fig6}
\end{figure}

In addition to determining the measurement efficiency we can also determine the transfer efficiency $\eta_{\mathrm{st}}$ independently. For a transfer field with a square envelope, we capture a coherent temporal mode that has an envelope \textit{growing} exponentially at a rate controlled by the strength of the transfer field. This fact is easily understood by imagining the capture process as the time reverse of the measurement process. In Fig. 6, we both demonstrate the shape of the captured mode and measure the transfer efficiency directly. Figure 6a depicts the protocol for capture, storage and retrieval where the preparation field is a growing exponential with rate $\Gamma_\mathrm{prep} = 2\pi \times 2.55$ kHz. The blue points in Fig. 6b show the percentage of retrieved coherent energy after being stored in the mechanical oscillator for 500 $\mu$s. The red points show the percentage of the preparation field that passes through the transmission line without being captured by the mechanics. The lines are predictions of the linear equations of motion that describe our system, with the system parameters determined independently. We find excellent agreement. As the percentage of recovered energy measured in Fig.~6 includes two transfer steps, a capture and a retrieval, $\% N_{\mathrm{coh}} \propto \eta_{\mathrm{st}}^2$. As expected, we find the maximum efficiency for a single transfer is $\eta_{\mathrm{st}} = 0.33$, when $\Gamma=\Gamma_\mathrm{prep}$ as shown by the green line in Fig. 6b. In the limit of a very large transfer field we see optomechanically induced transparency as the red points approach unity. By coupling the circuit to only one direction of the transmission line and by increasing the external coupling $\kappa_{\mathrm{ext}}$, we expect $\eta_{\mathrm{st}}$ to approach 1 in future experiments.

\section{\textbf{Thermal State Distribution}}

In Fig. 2a we plotted 2000 single shot quadrature measurements of the mechanical oscillator when it was in equilibrium at 25 mK. Here we show explicitly that the measured distribution for these points is consistent with a thermal distribution. Figure 7 shows a histogram of the inferred quanta (black points) for the data of Fig. 2a, a thermal state at 25 mK with $\eta =0.12$. Each bin in the histogram is five quanta wide. The error bars are one standard deviation statistical error derived from the number of counts in each bin. The red line shows the expected number of measurement outcomes yielding a particular $|\alpha|^2=X_1^2 + X_2^2$ based on a probability distribution function $P=(N_{\mathrm{th}}+1/\eta)^{-1}\exp(-|\alpha|^2(N_{\mathrm{th}}+1/\eta)^{-1})$, the distribution for a thermal state of occupancy $N_{\mathrm{th}}=50$ measured with efficiency $\eta$.

\begin{figure}
  \centering
  \includegraphics[width= 3 in]{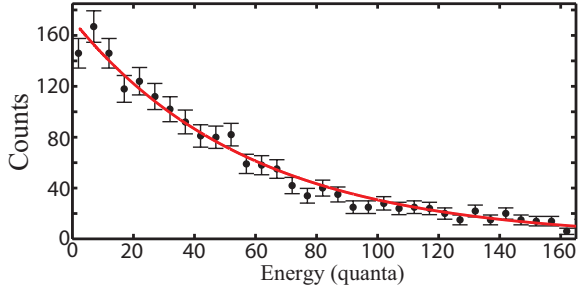}\\
  \caption{\textbf{Thermal state Distribution.}
  }\label{fig:fig7}
\end{figure}

\section{\textbf{State swaps}}
In Fig. 4, we demonstrated the ability to coherently transfer a state between the resonator and oscillator. By varying the length of the transfer pulse we were able to observe oscillations in $N_{\mathrm{coh}}$ as the state continuously swapped back and forth between the two systems. The solid line in Fig. 4b represents a fit to these oscillations. To derive the expression used in the fit, we start from the linearized Hamiltonian in the rotating frame of the transfer drive $\omega_\mathrm{d}$
\begin{equation}
H=-\hbar\Delta\hat{a}^\dagger\hat{a}+\hbar \Omega_\mathrm{m}\hat{b}^\dagger\hat{b}+\hbar g(\hat{a} \hat{b}^\dagger+ \hat{b} \hat{a}^\dagger) \;,
\end{equation}
where $\Delta=\omega_\mathrm{d}-\omega_\mathrm{c}$ and $g=g_\mathrm{0} \sqrt{N_\mathrm{d} (t)}$. For an optimally red-detuned drive, $\Delta=-\Omega_\mathrm{m}$. Using input-output formalism we write the linearized Heisenberg equations of motion \cite{Marquardt2007},

\begin{widetext}
\begin{equation}\label{eq:FullIO}
 \frac{d}{dt}
 \left[\begin{array}{c}\hat{a} \\ \hat{b}\\ \hat{a}^\dagger\\ \hat{b}^\dagger \end{array}\right]
 =
 \left[\begin{array}{cccc}i\Delta - \kappa/2 & -ig & 0 & -ig\\ -ig & -i\Omega-\Gamma_\mathrm{m}/2 & -ig & 0 \\ 0 & ig & -i\Delta-\kappa/2 & ig \\ig & 0 & ig & i\Omega-\Gamma_\mathrm{m}/2 \end{array}\right]
\left[\begin{array}{c}\hat{a} \\ \hat{b}\\ \hat{a}^\dagger\\ \hat{b}^\dagger \end{array}\right]+
\left[\begin{array}{cccc}-\sqrt{\kappa} & 0 & 0 & 0\\ 0 & -\sqrt{\Gamma_\mathrm{m}} & 0 & 0 \\ 0 & 0 & -\sqrt{\kappa} & 0 \\0 & 0 & 0 & -\sqrt{\Gamma_m} \end{array}\right]
\left[\begin{array}{c}\hat{\xi}_\mathrm{c} \\ \hat{\xi}_\mathrm{m}\\ \hat{\xi}_\mathrm{c}^\dagger\\ \hat{\xi}_\mathrm{m}^\dagger \end{array}\right]
\end{equation}
\end{widetext}

We introduce $\hat{\xi}_\mathrm{m}$ and $\hat{\xi}_\mathrm{c}$ as operators associated with the modes of the mechanical oscillator's environment and circuit's environment, respectively. Equation~\ref{eq:FullIO} can be transformed to our quadrature basis through

\begin{equation}
\left[\begin{array}{c}\hat{Z}_1 \\ \hat{X}_1 \\ \hat{Z}_2 \\ \hat{X}_2 \end{array}\right]
 =
\frac{1}{\sqrt{2}} \left[\begin{array}{cccc}1&0&1&0\\0&1&0&1\\-i&0&i&0\\0&-i&0&i \end{array}\right]
\left[\begin{array}{c} \hat{a} \\ \hat{b} \\ \hat{a}^\dagger \\ \hat{b}^\dagger \end{array}\right] \;,
\end{equation}
where $\hat{Z_1}$ and $\hat{Z_2}$ are the quadrature amplitudes of the resonator.

For the solid curve in Fig. 4b, we time evolve an initial coherent state $\langle \hat{X}_1 \rangle =6300$ for a constant coupling drive and calculate $N_{\mathrm{coh}}=\langle \hat{X}_1 \rangle^2+\langle \hat{X}_2 \rangle^2$ as a function of time. Because $\hat{\xi}_\mathrm{m}$ and $\hat{\xi}_\mathrm{c}$ assume random values with mean zero, they play no role in determining $N_{\mathrm{coh}}$.

\end{appendix}

\end{document}